\documentclass{jpsj-suppl}
\usepackage{txfonts} 

\title{Perturbative Analysis of Potential Scattering Problems in the Lieb-Liniger Model}

\author{Hironobu \textsc{Fujishima}$^{1}$ and Tetsu \textsc{Yajima}$^{2}$}

\inst{$^{1}$Optics R\&D Center, CANON INC., 23-10, Kiyohara-Kogyodanchi,
Utsunomiya 323-3298, Japan\\
$^{2}$JDepartment of Information Systems Science, Graduate School of
Engineering, Utsunomiya University, Yoto 7-1-2, Utsunomiya 321-8585,
Japan}

\email{H.Fujishima@gmail.com}

\recdate{July 15, 2013}

\abst{The Lieb-Liniger model which has a weak external potential term under the periodic boundary condition is investigated. By exploiting the Bethe states as bases, we perform a perturbation analysis up to the first order to obtain the shifts of eigenenergies and corresponding eigenstates which have been brought about by the external potential. If we take a sufficiently large system, the eigenstates can be ``the Schr\"odinger's cat states". Expectation values of the density operator taken between two Bethe states can be calculated with the aid of the Slavnov's formula and we evaluate the influence of the many-body interaction to the system under the external potential. The system is insensitive to the external potential because of the many-body interaction.}
\kword{the Lieb-Liniger model, external potential, the Schr\"odinger's cat state}

\begin{document}
\maketitle

\section{Introduction}
The Lieb-Liniger (LL) model is a well-known exactly solvable model which directly corresponds to a cold atom system confined in a very narrow potential\cite{Bloch,Lieb}. This model describes a one-dimensional system that consists of $N$-bosons interacting through the delta function potential. The Hamiltonian is expressed as
\begin{equation}
\hat{H}=\int_{0}^{L}\left(\frac{\partial \hat{\psi^{\dag}}(x)}{\partial x}\frac{\partial \hat{\psi}(x)}{\partial x}+c\hat{\psi}^{\dag}(x)\hat{\psi}^{\dag}(x)\hat{\psi}(x)\hat{\psi}(x)\right)dx,
\end{equation}
where we impose the periodic boundary condition with a period $L$. The field operators obey usual bosonic canonical commutation relations. The coupling constant is denoted by $c$. If $c$ tends to infinity, this model comes down to the Tonks-Girardeau Model. The total momentum operator is naturally defined as
\begin{equation}
\hat{P}=-i\int_{0}^{L}\hat{\psi^{\dag}}(x)\frac{\partial \hat{\psi}(x)}{\partial x}dx,
\end{equation}
and $[\hat{P},\hat{H}]=0$ holds. We can construct simultaneous eigenstates of the Hamiltonian and the momentum operator for an arbitrary value of $c$ by means of the Bethe ansatz Method\cite{Korepin} as
\begin{equation}
\hat{H}|P,E \rangle=\left(\sum_{j=1}^{N}k_{j}^{2}\right)|P,E \rangle,
\end{equation}
\begin{equation}
\hat{P}|P,E \rangle=\left(\sum_{j=1}^{N}k_{j}\right)|P,E \rangle=\left(\frac{2\pi}{L}\sum_{j=1}^{N}I_{j}\right)|P,E \rangle,
\end{equation}
where $I_{i}$'s are integers for odd $N$. In the case of $N$ being even, all $I_{i}$'s become half integers. The $k_{i}$'s are called rapidities and regarded as interaction-renormalized momentum of particles. They are determined by the Bethe equation,
\begin{equation}
k_{j}L=2\pi I_{j}-2\sum_{l\neq j}\mathrm{arctan}\left(\frac{k_{j}-k_{l}}{c}\right).
\end{equation}
Although the LL model fully deals with inter-particle interaction, it lacks the external potential term which is an indispensable part of our main concern. In this paper, we shall add a weak external potential term to the original LL model. Our Hamiltonian is written with small parameter $\epsilon$ as 
\begin{equation}
\hat{H}=\int_{0}^{L}\left(\frac{\partial \hat{\psi^{\dag}}(x)}{\partial x}\frac{\partial \hat{\psi}(x)}{\partial x}+c\hat{\psi}^{\dag}(x)\hat{\psi}^{\dag}(x)\hat{\psi}(x)\hat{\psi}(x)+\epsilon V(x)\hat{\psi}^{\dag}(x)\hat{\psi}(x)\right)dx.
\end{equation}
Although the exact integrability is destroyed by adding the potential term, utilizing the complete orthonormal set of bases established above, we can perturbatively investigate the change of eigenenergies and eigenstates.
\section{First Order Perturbation Theory and the Slavnov's Formula}
In this section, we summarize the result of the first order perturbation theory and explain the Slavnov's Formula. We shall only consider non-degenerated cases since our non-perturbed state is non-degenerated as long as we treat the ground state. Our main concerns in this work are the followings:
In comparison to the ordinary one-particle quantum mechanics (QM), 

1)How the many-body effect alters and affects the result of QM calculation.

2)The way the many-body effect appears in the final expressions.

To the first order of $\epsilon$, the eigenenergy changes to
\begin{equation}
E'=E^{(0)}+\epsilon E^{(1)}=\sum_{j=1}^{N}k_{j}^{2}+\epsilon\langle P,E|\left(\int_{0}^{L}V(x)\hat{\psi}^{\dag}(x)\hat{\psi}(x)dx\right)|P,E\rangle.
\end{equation}
Using the translation relations  $\hat{\psi}(x)=e^{i\hat{P}x}\hat{\psi}^{\dag}(0)e^{-i\hat{P}x}$, we get
\begin{equation}
E'=\sum_{j=1}^{N}k_{j}^{2}+\epsilon\langle P,E|\hat{\psi}^{\dag}(0)\hat{\psi}(0)|P,E\rangle\int_{0}^{L}V(x)dx.\label{1}
\end{equation}
In the similar way, the corresponding eigenstate of the total Hamiltonian is calculated as
\begin{equation}
|P,E\rangle '=|P,E\rangle+\epsilon\sum_{E'\neq E}\frac{\langle P',E'|\hat{\psi}^{\dag}(0)\hat{\psi}(0)|P,E\rangle}{E-E'}|P',E'\rangle\int_{0}^{L}V(x)e^{-i(P-P')x}dx.\label{key}
\end{equation}
In these expressions, the terms like $\langle P',E'|\hat{\psi}^{\dag}(0)\hat{\psi}(0)|P,E\rangle$ are called form factors and exactly given by the Slavnov's formula\cite{Gaudin,Slavnov,Kojima,Caux,Sato}. Especially, for the case $P=P'$, the form factor is equal to unity. If $P$ differs from $P'$, the corresponding form factor is less than unity and calculated in a rather complicated way. We find
\begin{equation}
\langle P',E'|\hat{\psi}^{\dag}(0)\hat{\psi}(0)|P,E\rangle=(-1)^{\frac{N(N+1)}{2}}(P-P')\left(\prod_{j,l=1}^{N}\frac{1}{k'_{j}-k_{l}}\right)\left(\prod_{j>l}^{N}k_{j,l}k'_{j,l}\sqrt{\frac{K(k'_{j,l})}{K(k_{j,l})}}\right)\frac{\mathrm{det}U(k,k')}{\sqrt{\mathrm{det}G(k)\mathrm{det}G(k')}},
\end{equation}
where, $k_{j,l}=k_{j}-k_{l}$ and $k'_{j,l}=k'_{j}-k'_{l}$. The kernel $K(k)$ is defined as $K(k)=2c/(k^2+c^2)$. The Gaudin matrix $G(k)$ is given as $G(k)_{j,l}=\delta_{j,l}[L+\sum_{m=1}^{N}K(k_{j,m})]-K(k_{j,l})$. The matrix $U(k,k')$ is of $(N-1)$ by $(N-1)$ one, whose $(j,l)$-th element is
\begin{equation}
U(k,k')_{j,l}=2\delta_{j,l}\mathrm{Im}\left[\prod_{a=1}^{N}\frac{k'_{a}-k_{j}+ic}{k_{a}-k_{j}+ic}\right]+\frac{\prod_{a=1}^{N}(k'_{a}-k_{j})}{\prod_{a\neq j}^{N}(k_{a}-k_{j})}\times(K(k_{j,l})-K(k_{N,l})).
\end{equation}
Expressions (\ref{1}) and (\ref{key}) directly answer our questions 1)-2) stated before. They might look simple, however, at least they show the following non-trivial facts. A) The shift of the eigenenergies is almost the same as the QM case and the many-body effect does not appear explicitly. B) On the contrast to  the shift of the eigenenergies, the many-body interaction clearly alters the result of QM when we consider new eigenstates. The many-body effect appears in the expression (\ref{key}) as suppression factors for each superposed eigenstates. C) The suppression factors mentioned above are all exactly calculable quantities thanks to the Slavnov's formula. These non-trivial facts have their own theoretical novelty and importance since they enable us making predictions on observable quantities as shown in the following sections.
\section{New Ground State of the System Applied a Weak Sinusoidal Potential}
In this section, we consider the first order change of the eigenstates. Here, we limit our consideration to the ground state $|P,G\rangle=|0,E_{G}\rangle$, which is not degenerated. 

Let us consider a sinusoidal potential $V(x)=\mathrm{cos}(\frac{2\pi n}{L}x)$, where $n$ is an integer. Since any periodic potentials on the one-dimensional ring can be expressed by a linear superposition of the sinusoidal functions with various values of $n$, studies on the sinusoidal potential have fundamental importance. Note that this potential makes $E^{(1)}=0$. Because the integral part of (\ref{key}) gives
\begin{equation}
\int_{0}^{L}V(x)e^{iPx}dx=\frac{L}{2}(\delta_{P=\frac{2\pi n}{L}}+\delta_{P=\frac{-2\pi n}{L}}),
\end{equation}
only states that satisfies $|P|=\frac{2\pi n}{L}$ contribute to the summation in (\ref{key}). In general, infinitely many states share the same $n$ because there exist infinitely many excited states. If we label them by a letter $m$, the new ground state becomes 
\begin{equation}
|0,E_{G}\rangle+\frac{\epsilon L}{2}\sum_{m\neq G}\left(\frac{\langle \frac{2\pi n}{L},E_{m}|\hat{\psi}^{\dag}(0)\hat{\psi}(0)|0,E_{G}\rangle}{E_{G}-E_{m}}|\frac{2\pi n}{L},E_{m}\rangle+\frac{\langle -\frac{2\pi n}{L},E_{m}|\hat{\psi}^{\dag}(0)\hat{\psi}(0)|0,E_{G}\rangle}{E_{G}-E_{m}}|-\frac{2\pi n}{L},E_{m}\rangle\right).\label{G}
\end{equation}
This is nothing except a quantum superposition of macroscopically distinct states, or {\it the Schr\"odinger's cat state}, since our states can describes systems with sufficiently large $L$ and $N$. 

The superposed states are corresponding to two opposite directions of superfluid flow of the quantum gas. We can calculate the density profile of the quantum gas on which a weak external potential $\epsilon V(x)=\epsilon\mathrm{cos}(\frac{2\pi n}{L}x)$ is applied. For example, by taking the expectation value of the density operator $\hat{\rho}(x)=\hat{\psi}^{\dag}(x)\hat{\psi}(x)$ regarding the new ground state (\ref{G}), we get    
\begin{equation}
\rho(x)=1+2\epsilon \sum_{m\neq G}\left(\frac{|\langle \frac{2\pi n}{L},E_{m}|\hat{\psi}^{\dag}(0)\hat{\psi}(0)|0,E_{G}\rangle|^2}{E_{G}-E_{m}}\right)\mathrm{cos}(\frac{2\pi n}{L}x),\label{fr}
\end{equation}
to the first order of $\epsilon$.
\section{Numerical Estimation of the Compliance Factor}
Among our results, the expression (\ref{fr}) has particular significance. The coefficient in front of cosine
\begin{equation}
\sum_{m\neq G}\left(\frac{|\langle \frac{2\pi n}{L},E_{m}|\hat{\psi}^{\dag}(0)\hat{\psi}(0)|0,E_{G}\rangle|^2}{E_{G}-E_{m}}\right)\label{cf}
\end{equation}
 makes quantitative prediction on how compliant the system responds to the external force. Since this quantity is related to the density profile of the system, it is directly observable and can be the subject to experimental verification. We name it the compliance factor (CF) and  make a numerical estimation of this value for some parameters. 

Firstly, we consider the external potential that has only one node, namely $n=\sum_{i=1}^{N}I_{i}=1$. Secondly, we try the double-node case, $n=2$. We simultaneously vary the number of particles $N$ and the periodic system size $L$ from 5 to 11, keeping the particle number density without the external potential $N/L$ unity. Therefore, the dimensionless parameter  $\gamma=\frac{cL}{N}$  that determines the property of the LL gas is coincident with the coupling constant $c$. The range of $c$ is $-15\leq\mathrm{log_{2}}c=\mathrm{lb}c\leq10$. The suffix $m$ indicates that we sum up all the excited states that share the same value of $n$ except the ground state. Although there exist infinite excited states, this series coverages quite quickly since the contributions from highly excited states with large eigenenergies are very small and can be neglected.
\begin{figure}[hh]
\begin{minipage}{0.5\hsize}
\begin{center}
\includegraphics[clip,width=75mm]{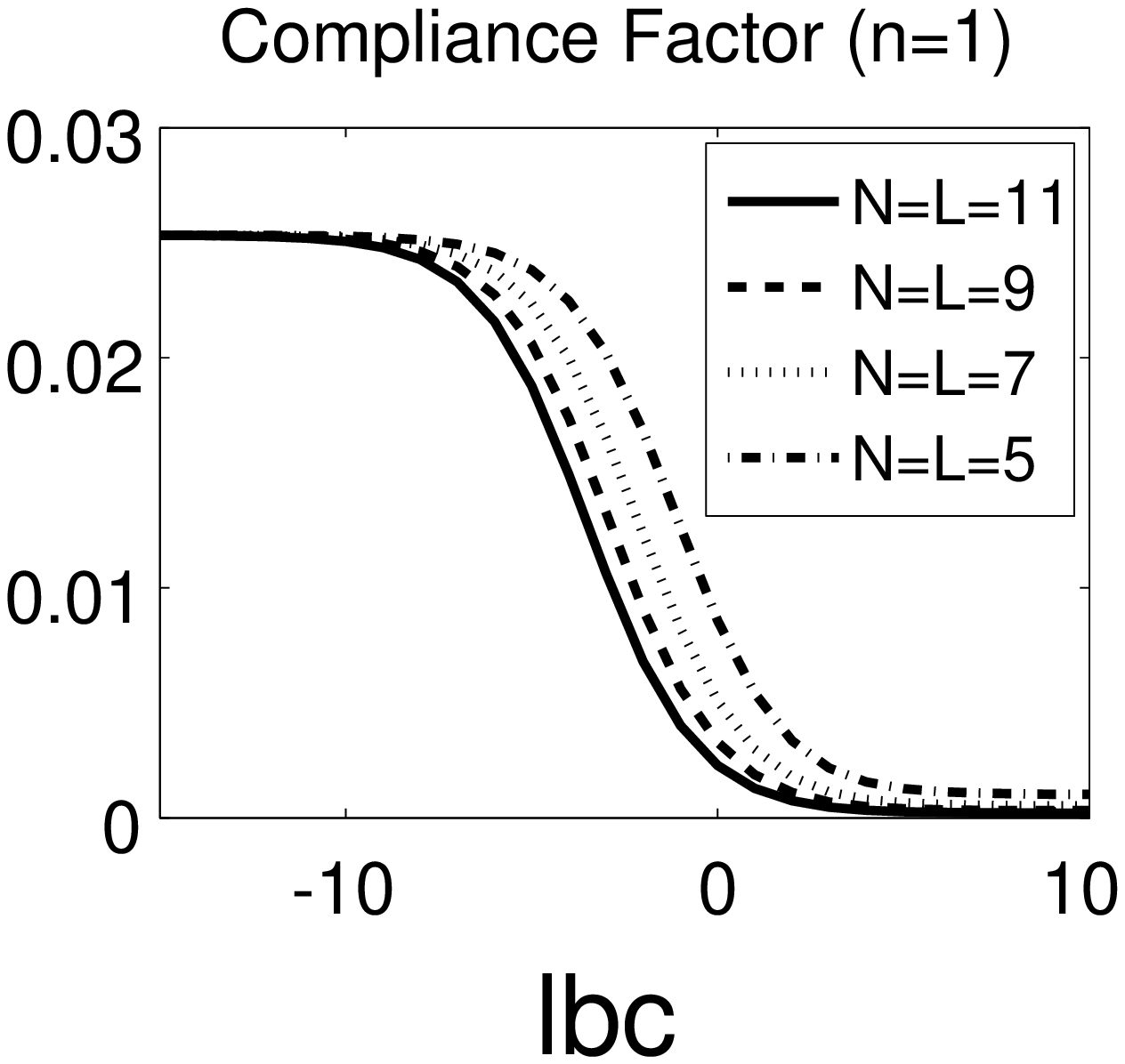}
\end{center}
\end{minipage}
\begin{minipage}{0.5\hsize}
\begin{center}
\includegraphics[clip,width=68mm,height=72mm]{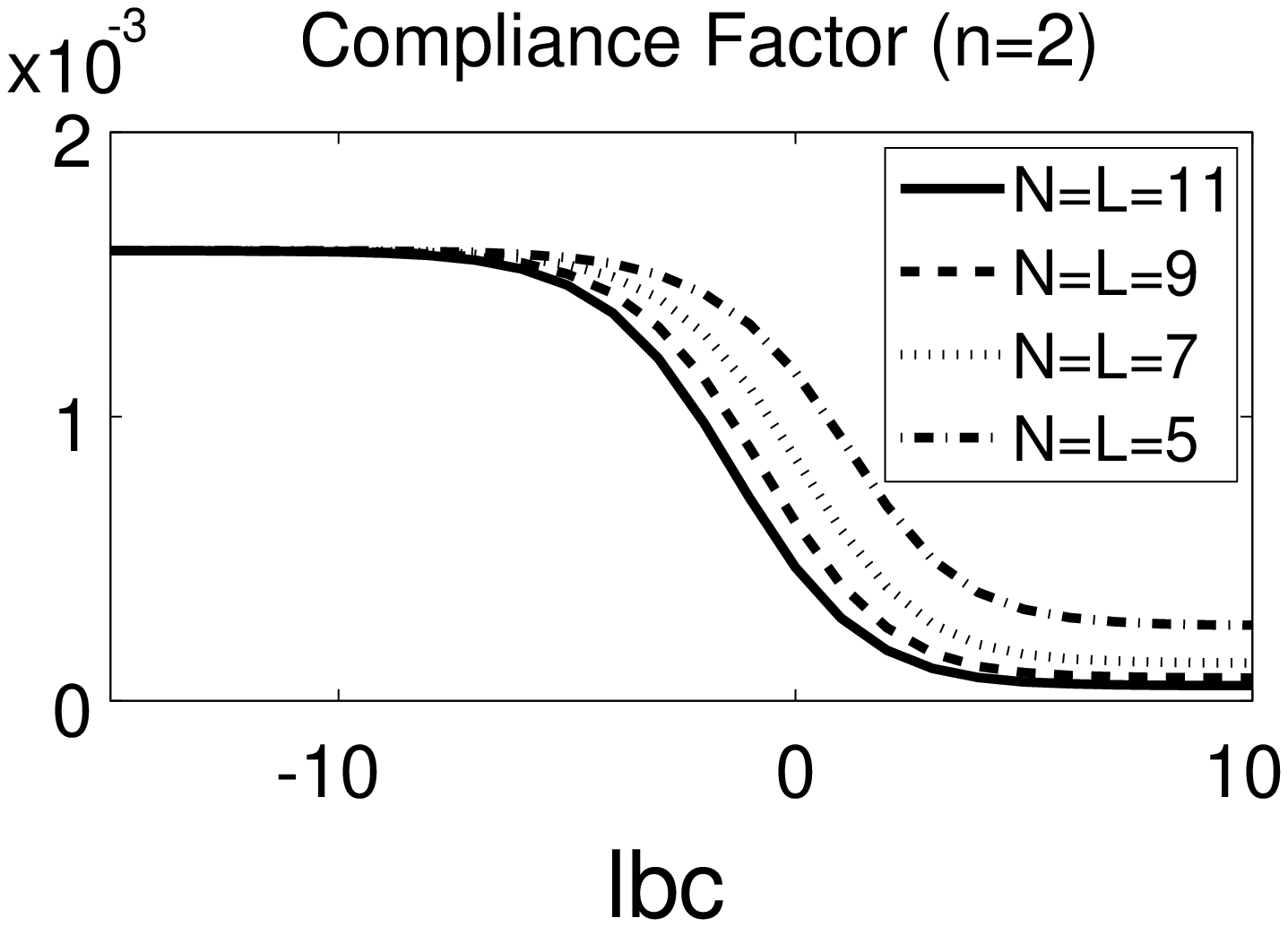}
\end{center}
\end{minipage}
\end{figure}
From this calculation, we have lead following conclusions.

1)The CF monotonically decreases as the coupling constant $c$ grows. The gradients of decrease get smaller for smaller particle number cases. 

2)For larger $n$, the CF takes very small values and  starts to fall down at larger value of $c$. 

3) At the both limits $c\to 0$ and $c\to\infty$, the CF converges to respective non zero values.

4)The finite size effect for smaller $N$ and $L$ cases manifests itself as the larger limiting value of the CF at $c\to\infty$. Moreover, the dropping points of the CF are shifted to larger values of $c$.   
\section{Experimental Realization}
To observe the results in the previous section experimentary, the first candidate system described by our modified version of the LL model is the system of cold bosons confined in a one-dimensional trap. Such a system including the Tonks-Girardeau gas has already been realized\cite{Weiss,Paredes}. Actually, the traps used for the experiments were straight and not of ring- or torus-like. The truly ring or torus-like trap was created in 2011\cite{Cambell}. External potentials are available by various methods though generating the sinusoidal potential along the ring might be challenging. Therefore, there exist ample possibilities for experimental realization of the system we have considered in this paper. 
\section{Summary and Future Work}
We have investigated the LL model with a weak external potential by the perturbation method. We wrote down the shifts of the ground energy and corresponding ground state in terms of the Bathe states up to the first order. If we take a sufficiently large system, the ground state can be ``the Schr\"odinger's cat state".  The many-body effect appeared in expectation values of the density operator taken between two Bethe states, which can be calculated using the Slavnov's formula. We found the many-body interaction made the system insensitive to the external potential. We made some comments on prospective experiments of our model.
 
 For a strongly interacting regime, it is reported that the system underwent ``the pinning transition" and an arbitrarily weak optical lattice led to immediate pinning of the atoms, provided $N\sim n$ holds\cite{Haller}. This pure quantum phase transition is described by ``quantum sine-Gordon model", which is another exactly solvable quantum field theoretical model derived as an effective theory for the system we have considered. The relationship between our LL model with a weak external potential and the quantum sine-Gordon model might be a fascinating subject of future works. 
\section*{Acknowledgment}
One of the authors, H. F. thanks Utsunomiya University for the hospitality and for offering wonderful working spaces and
opportunities of fruitful discussion.


\begin{thebibliography}{12}
\bibitem{Bloch} I. Bloch, J. Dalibard, and W. Zwerger: Rev. Mod. Phys. \textbf{80} (2008) 885.
\bibitem{Lieb} E. H. Lieb and W. Liniger: Phys. Rev. \textbf{130} (1963) 1605; E. H. Lieb: Phys. Rev. \textbf{130} (1963) 1616.
\bibitem{Korepin} V.E. Korepin, N.M. Bogoliubov, and A.G. Izergin: {\it Quantum Inverse Scattering Method and Correlation Functions} (Cambridge University Press, Cambridge, 1993).
\bibitem{Gaudin} M. Gaudin: {\it La fonction d'onde de Bethe} (Masson, Paris, 1983); V. E. Korepin: Commun. Math. Phys. \textbf{86} (1982) 391.
\bibitem{Slavnov} N. A. Slavnov: Teor. Mat. Fiz. \textbf{79} (1989) 232; \textbf{82}, 389 (1990).
\bibitem{Caux} J.-S. Caux, P. Calabrese, and N. A. Slavnov: J. Stat. Mech. (2007) P01008.
\bibitem{Kojima} T. Kojima, V. E. Korepin, and N. A. Slavnov: Commun. Math. Phys. {\bfseries 188} (1997) 657.
\bibitem{Sato} J. Sato, R. Kanamoto, E. Kaminishi, and T. Deguchi: Phys. Rev. Lett. \textbf{108} (2012) 110401.   
\bibitem{Weiss} Toshiya Kinoshita, Trevor Wenger, and David S. Weiss: Science {\bf 305} (2004) 1125.
\bibitem{Paredes} B. Paredes, A. Widera, V. Murg, O. Mandel, S. F\"olling, I. Cirac, G. V. Shlyapnikov, T. W. H\"ansch, and I. Bloch: Nature {\bf 429} (2004) 277.
\bibitem{Cambell} A. Ramanathan, K. C. Wright, S. R. Muniz, M. Zelan, W. T. Hill, III, C. J. Lobb, K. Helmerson, W. D. Phillips, and G. K. Campbell: Phys. Rev. Lett. \textbf{106} (2011) 130401.
\bibitem{Haller} E. Haller, R. Hart, M.J. Mark, J.G. Danzl, L. Reichsollner, M. Gustavsson, M. Dalmonte, G. Pupillo, and H.-C. Nagerl: Nature {\bfseries66} (2010) 597.
\end{thebibliography}
\end{document}